**Atomistic and continuum modeling of non-equilibrium melting of aluminum**

Alexander E. Mayer[1], Vasiliy S. Krasnikov

*Chelyabinsk State University, Bratiev Kashirinykh 129, Chelyabinsk 454001, Russia*

**Abstract**

MD simulations of the non-equilibrium melting of aluminum are performed both with and without accounting of the electronic subsystem. A continuum model of melting is purposed basing on the obtained MD results, in which the current phase state is described in terms of fields of concentration and size of melting sites. Growth equation for melting areas is derived from the heat fluxes analysis. Nucleation equation for melting sites is formulated basing on the thermofluctuational approach. The method of determination of the model coefficients with using the MD simulation results is purposed. The continuum model is applied to the problem of the non-equilibrium melting of aluminum within the energy absorption area of the high-current electron beam.

---

[1] Corresponding author. Tel.: +7(351)799-71-61; fax: +7(351)742-09-25; e-mail: mayer@csu.ru, mayer.al.evg@gmail.com

## 1. Introduction

Metastable state of an overheated solid metal can be realized in the case of rapid temperature increase [1]; this state decays in the course of time by means of complete or partial melting. Melting at such condition is considerably non-equilibrium and proceeds with a finite rate. Metal irradiation by powerful laser [2-4] or high-current electron beam [5], as well as compression in strong shock wave [6,7], are the typical situations, in which the non-equilibrium melting takes place. Large-scale molecular dynamics (MD) simulation [2,3,6-8] is the direct method of numerical investigation of this phenomenon. MD is applicable for the problem of metal exposure to an ultra-short powerful laser pulse and for the problem of propagation of a shock wave with a narrow front when the evolution of phase state occurs in a thin layer of matter during short time interval. Direct application of MD for most of other problems, for instance, description of melting under the action of high-current electron irradiation, is, at least, inefficient approach due to large spatial and time scales of these problems. In the case of high-current electron irradiation, the spatial scale reaches up to several millimeters, and the time scales varies from tens of nanoseconds up to several microseconds. In present work we propose a continuum model of non-equilibrium melting of aluminum basing on our MD investigations and apply this model for description of the matter behavior in the energy absorption area of the high-current electron beam.

Besides the direct modeling of the physical processes accompanied by melting, the model problem statements are widely used in atomistic simulations for determination of various characteristics of the solid-melt phase transition, such as the kinetics coefficient [9-12], which controls the propagation velocity of the interface between phases, the nucleation rate of melting or crystallization sites [13,14]. The obtained results are used, for example, for construction of the phase field model [10,11], in which all interfaces are traced through solution of field equations. The continuum model proposed by us is based on the description of the current phase state by means of fields of concentration and size of the melting sites; the site size is supposed to be much less than the typical spatial scale of the considered problem; so, we do not trace all interfaces explicitly, but view an averaged picture instead. Two model problem statements are used by us for determination of the model coefficients: (i) the reaching of equilibrium in an isolated system with a flat interface between solid and liquid; (ii) the melting of an initially homogeneous monocrystal heated by a source with constant power of energy release. It is well-known that the local temperature distribution near the interphase boundary [10-12] and the electronic thermal conductivity [15] considerably influence the propagation velocity of the boundary. Our analysis shows that the changeover of the material structure requires a negligible time during the melting or crystallization front propagation, while the propagation velocity is completely determined by the heat fluxes near the front, it means, by the thermal conductivity.

## 2. MD setup

Atomistic simulations of melting kinetics is performed with the help of LAMMPS package [16], part of the calculations uses the GPU library [17] and runs on video cards. Interatomic force field is described by an EAM potential, which was especially developed to accurately reproduce the phase transition solid-liquid in aluminum [18]. Since heat transfer in metals is carried out largely by electrons, the package LAMMPS is compiled with TTM library realizing two temperature model of heat transfer through the electron subsystem in calculation area [19,20]. The parameters of the electronic thermal conductivity model are taken from [21,22] for temperature of 1000 K: heat capacity of electronic gas $C_e = 1.3 \cdot 10^5$ $J \cdot m^{-3} \cdot K^{-1}$, heat conductivity coefficient $\kappa_e = 110$ $W \cdot m^{-1} \cdot K^{-1}$, intensity of heat exchange between ions and electrons $g_p = 3 \cdot 10^{17}$ $W \cdot m^{-3} \cdot K^{-1}$. We use the OVITO [23] tool for visualization of the obtained atom distributions and the centrosymmetrical parameter [24] to distinguish the solid and melt parts in these distributions.

The melting curve of aluminum and the dependence the melting kinetics on the pressure and the rate of thermal energy supply are investigated via two formulations of the problem. In the first one, the movement of the phase solid-liquid boundary is simulated in a region including both phases initially at the same temperature. In this series of calculations we use a rectangular area with the size 40x20x20 nm that corresponds to one million of atoms. The periodic conditions are set at the boundaries of the area. The heat conduction equation is solved using the finite difference method [19,20] on the computational grid with discretization onto 100 cells along Ox axis and onto 10 cells along Oy and Oz axes. This choice of the spatial discretization of grid is associated with the feature of a plane crystallization front movement problem. The system is maintained at a constant temperature of 900 K and a given pressure during 5 ps. After that the heating of central part of the area between 10 and 30 nm along axis Ox is carried out at a constant volume up to the temperature of substance melting; and the heating is continued during 5 ps. Further, the temperature of the central part is dropped to some temperature $T_1$, for which the substance is kept in a liquid state; the stage cooling is lasted 30 ps. The temperature of the ambient substance is artificially set to constant value of $T_1$ during the heating-cooling cycle. Then, the system is maintained in the state with constant volume and energy until the average temperature of the whole area $T_2$ reaches the stationary level. The steady temperature and pressure in the system are further used as one point on the melting curve of aluminum.

The second series of calculations is aimed at the study of the kinetics of melting at different rates of energy supply. In this case, the computational area is a cube with sides of 24.5 nm, and there are 846 000 atoms in the area. In order to solve the heat conduction equation the grid with the same discretization along all axes of the coordinate system onto 60 cells is applied to the area, such choice of a spatial grid is caused by the fact that the forming melting sites have dimensions comparable with the interatomic distance. The system is preliminarily kept at a constant temperature of 900 K and a constant zero pressure during 20 ps. After that, the system is heated at a constant pressure with a constant rate of energy supply, which is added to the atoms uniformly throughout the volume of the computational area.

## 3. Results of MD

The volume of melt can decrease or increase depending on the initial temperature and pressure of the area in the simulations of flat solid-liquid phase front movement. The volume of central melted area decreases (see Fig. 1), otherwise, the crystallization of substance occurs, if the initial average temperature of the area $T_1$ is smaller than the melting temperature of aluminum at the given pressure. The phase boundary is kept its flat form with good accuracy, at the same time, the asperities and cavities not exceeding a height of 3-4 interatomic distances are formed at the boundary, its positions vary randomly along the section perpendicular to the Ox axis. Over time, the average temperature in the area reaches a steady value $T_2$, after reaching of which the phase boundary stops the translational motion. The asperities and cavities are still randomly formed on its surface, but its average position remains unchanged (see Fig. 1 (200 and 300 ps)). The average temperature of the whole area increases during the movement of phase front due to the release of latent heat of melting in the crystallizing layer near the phase boundary (Fig. 2). On the contrary, the average temperature decreases if the initial temperature $T_1$ is higher than the melting point. But the same pressure in the systems provokes the same steady temperatures $T_2$ both for cases when the initial temperature is above the melting point and for systems with the initial temperature lower than the melting point.

Movement the solidification front deeper into the melt leads the release of latent heat of melting in a thin layer of atoms near the phase interface, as a result the atoms of the solid phase lying near the phase boundary have increased the average kinetic energy in comparison with other atoms of solids remote from the interface. Then the rate of heat transfer from the thin layer of atoms deep into the solid phase is to influence the speed of the front movement. We perform the additional calculations without electronic thermal conductivity to study the effect of thermal conductivity onto the establishing of the final temperature $T_2$. When the heat transfers exclusively by ionic subsystem of crystal, the rate of heat conductivity is substantially reduced that leads to a long establishing of the final steady temperature in the computational area (Fig. 2). Durations of steady temperature achievement can differ by 50-60% for the cases of calculations with and without electronic heat conductivity.

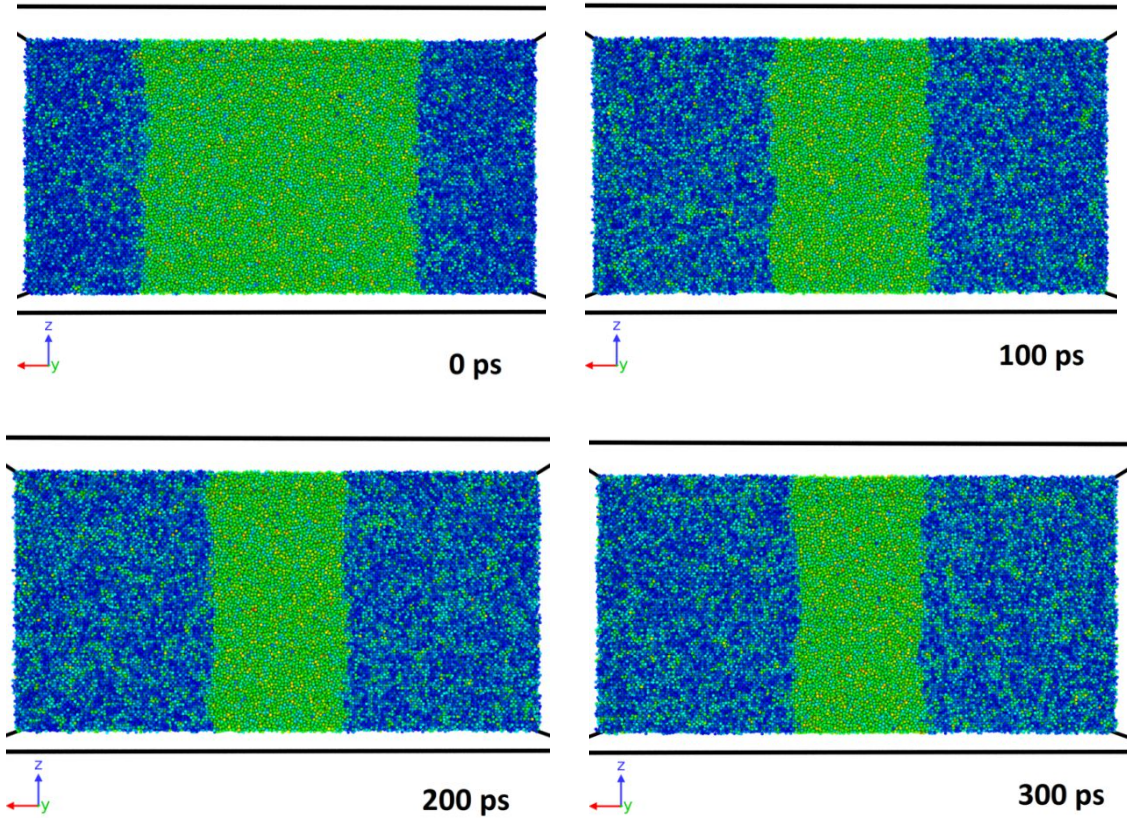

**Fig. 1.** The position of the phase solid-liquid boundary plane versus time. Distribution of centro-symmetry parameter: solid corresponds to the area mainly painted in blue (dark), melt – green (light).

Obtained from MD calculations value of melting temperature at a pressure of 0.2 GPa is equal to 933 K, which corresponds well with the melting point of aluminum at zero pressure. The melting temperature increases together with the rise of pressure, this dependence is presented in Fig. 3 and it can be approximated by the following function

$$T_{\mathrm{m}} = 933\,\mathrm{K} + 46.9\,\mathrm{K} \times \left(P/1\mathrm{GPa}\right) - 1.3\,\mathrm{K} \times \left(P/1\mathrm{GPa}\right)^{2} + 0.012\,\mathrm{K} \times \left(P/1\mathrm{GPa}\right)^{3}. \quad (1)$$

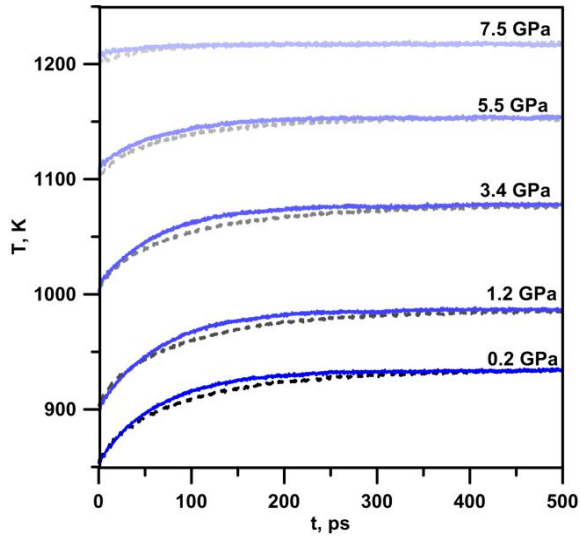

**Fig. 2.** Reaching the phase equilibrium in the region with the molten layer: the solid lines show the results of calculations with electronic heat conductivity, dashed - without.

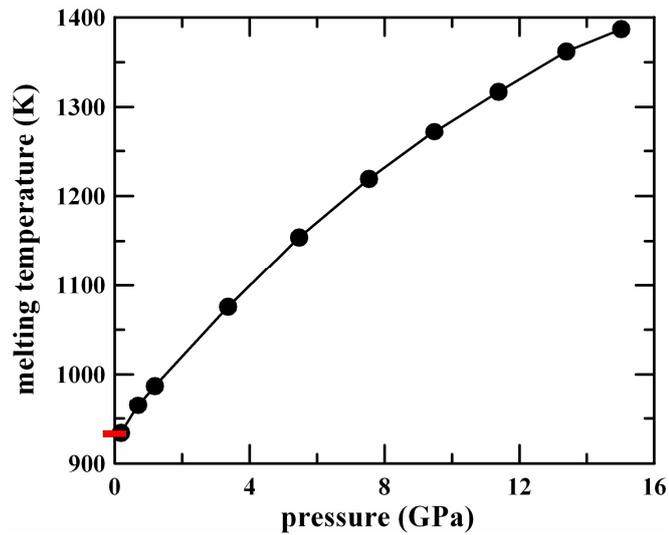

**Fig. 3.** Melting curve for aluminum obtained with MD calculations. Red dash is experimental value of the melting temperature at atmospheric pressure.

The supply of energy with a constant rate initially leads to an almost linear increase in temperature, which slope is determined by the heat capacity of the system (Fig. 5). A slight deviation from the linearity appears to be associated with the formation of nuclei of the liquid phase, but at the initial stage formation of critical nuclei does not occur (Fig. 4 (80 ps)). The sizes of the formed regions of the liquid phase rise with increase in temperature, and at some moment the formation of a critical nucleus befalls, which later does not disappear and acts as a stable center of the substance melting (Fig. 4 (95, 110 ps)).

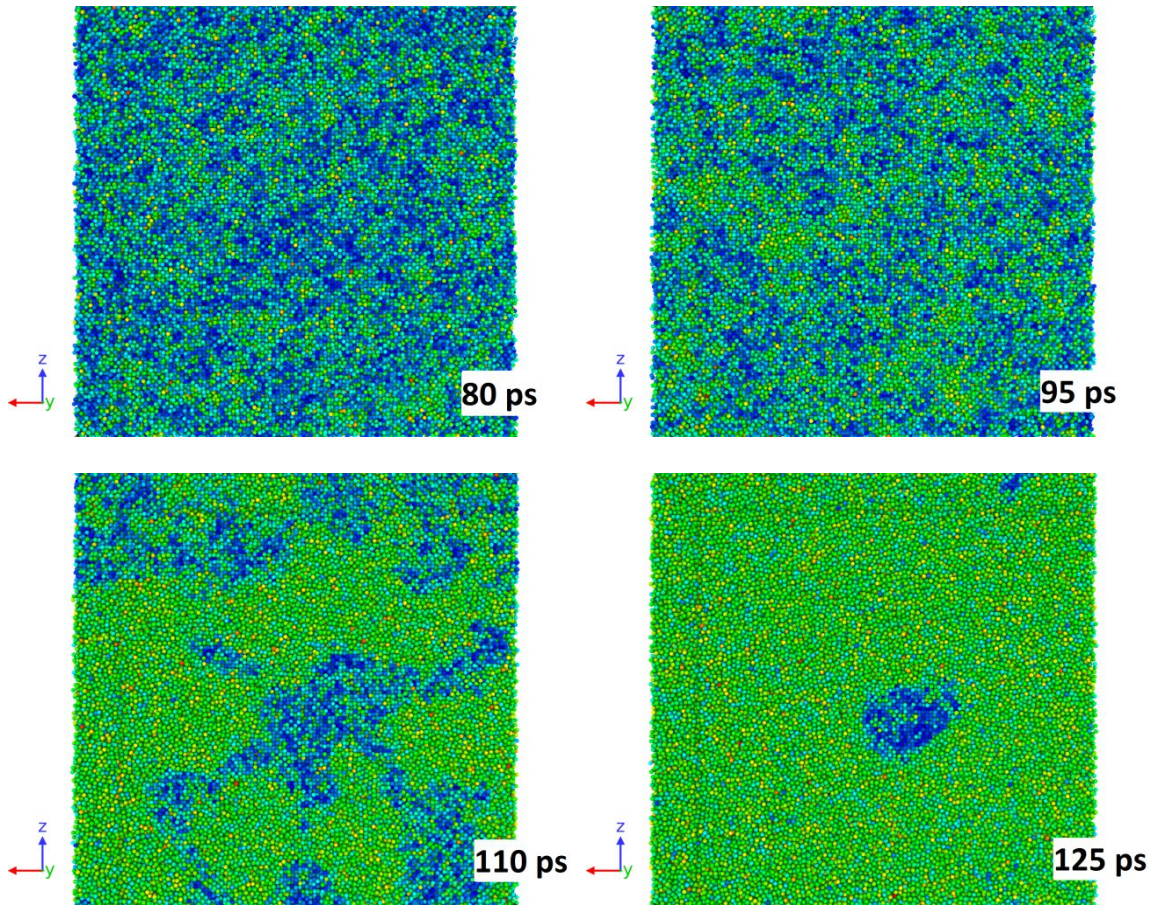

**Fig. 4.** Formation and growth of the nuclei of the melt in the solid phase versus time. The heating rate is 3.52 PW/kg. Solid corresponds to the area mainly painted in blue, melt - green.

Formation of critical nuclei and begun melting of substance leads to sharp drop of average temperature in area due to the conversion of thermal energy of the atoms in the latent heat of fusion (Fig. 5). After the complete melting of the substance in area the increase in temperature is resumed (Fig. 5). Rise of the rate of heat supply leads to the increase in the maximum temperature reached before the complete melting and to the lowering of the minimum in the temperature dependences because the heat supply partially covers the energy required for the melting.

Initial curve slope is somewhat smaller in the case of calculations with the electronic heat conductivity, which is connected with the contribution of the heat capacity of electrons. The maximal overheating is approximately the same for both types of calculations. A considerable difference takes place near the bottom of the energy drop, where the MD system without electronic subsystem demonstrates lower temperature. Analysis of the mean density evolution reveals that the complete melting occurs earlier in this case (without accounting of electrons), which is reflected on the temperature behavior. The later completion of melting for system with electrons can be caused by the fact that the electronic thermal conductivity distributes the excess thermal energy over the all volume of the system. In the case of absence of the electronic thermal conductivity, this thermal energy is concentrated inside the solid phase between the melting sites that accelerates the melting.

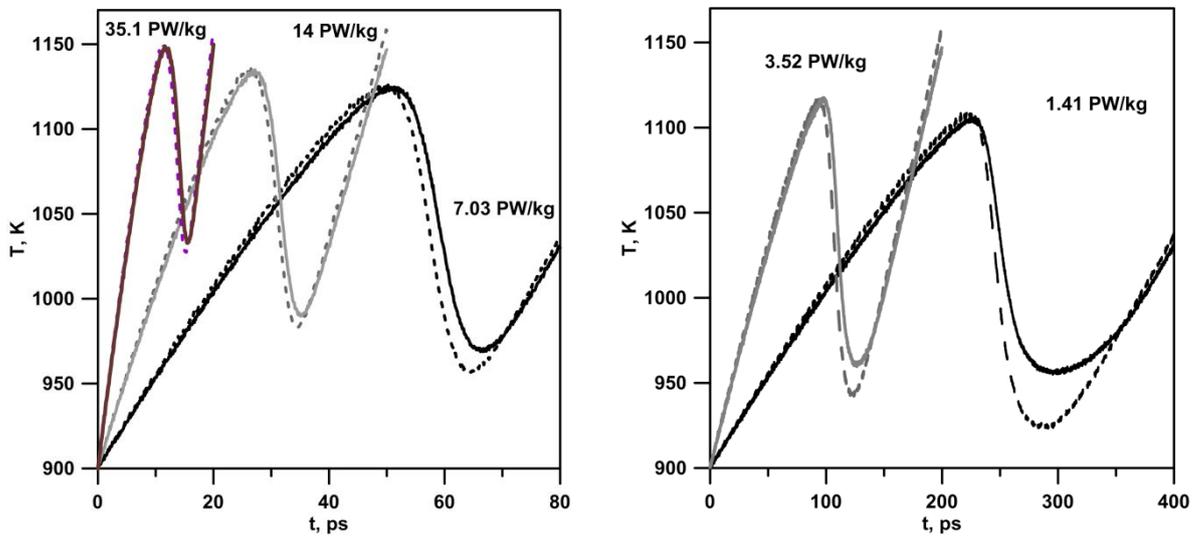

**Fig. 5.** The time profiles of the temperature of aluminum at a constant rate of energy supply. The solid and dashed lines correspond to the calculations with and without electronic thermal conductivity respectively.

## 4. Continuum model of melting

### 4.1. Propagation velocity of a flat melting/crystallization front

Suppose that the changeover of the material structure requires a negligible time during the melting or crystallization front propagation, while the propagation velocity is completely controlled by the thermal conductivity. Consider a plane layer of melt with the thickness of $2R$, which border with two identical flat layers of solid metal with the thickness of $L/2 - R$ each of them. The total thickness of system is $L$; the volume fraction of melt is $\alpha = 2R/L$. Zero heat fluxes are set on the external boundaries of the system–at $x = 0$ and $x = L$, where $Ox$ axis is normal to the interface between solid and melt; this problem statement corresponds to the periodic boundary conditions used in MD for tracing the flat front propagation (Section 2). Assume that the front propagation velocity is completely determined by the rate of the heat supply to the interface from the overheated solid layers. In this approximation, the balance between the heat flux and the rate of transition into the latent heat of melting gives us

$$\lambda \frac{\mathrm{d}R}{\mathrm{d}t} = c_V \chi \left( \nabla T \right)_R, \tag{2}$$

where $\lambda$ is the latent heat of melting; $c_V$ is the specific heat capacity; $\chi$ is the thermal diffusivity; $\left( \nabla T \right)_R$ is the temperature gradient in solid metal near the melt interface. Suppose that the temperature inside the molten layer is uniform and equal to the melting temperature $T_\mathrm{m}$. Stationary solution of the heat conduction equation gives one a linear increase of temperature from the value of $T_\mathrm{m}$ on the melting front up to the value $T_\mathrm{max}$ at the external boundaries of the system. The following expression can be written for the temperature gradient

$$\left( \nabla T \right)_R = \frac{T_\mathrm{max} - T_\mathrm{m}}{L/2 - R} = \frac{4}{L} \left( T - T_\mathrm{m} \right), \tag{3}$$

where $T$ is the volume-averaged temperature of the system. Combination of Eqs. (2) and (3) gives us the following equation for the melting front propagation velocity

$$\frac{\mathrm{d}R}{\mathrm{d}t} = \chi \frac{4}{L} \left( \frac{c_V}{\lambda} \right) \left( T - T_\mathrm{m} \right), \tag{4}$$

and equation for rate of change of the melt volume fraction

$$\frac{\mathrm{d}\alpha}{\mathrm{d}t} = \chi \frac{8}{L^2} \left( \frac{c_V}{\lambda} \right) \left( T - T_\mathrm{m} \right). \tag{5}$$

Eqs. (4) and (5) are derived within the quasi-stationary approximation, it means, the heat fluxes in the system steady much faster than the values of $R$ and $\alpha$ change. Similar equations can be derived for $R$ and $\alpha$ in the case of crystallization front if one suppose that temperature in the solid layers is uniform and equal to $T_m$, and the temperature gradient takes place inside the overcooled melt ($T < T_m$). The invariance of Eqs. (4) and (5) with respect to replacement of melting on crystallization is clear from the fact that the value $R$ is absent in the right-hand parts of these equations, which contain only $L$ –the total thickness or period of the system.

Phase transition changes the mean temperature of the system. In the case of absence of the external energy deposition and the work over system (this is the case of MD simulations for flat front) the temperature evolution is determined by the simple balance

$$\frac{dT}{dt} = -\left(\frac{\lambda}{c_V}\right)\frac{d\alpha}{dt} = \frac{8\chi}{L^2}\left(T - T_m\right). \qquad (6)$$

Integration of Eq. (6) over time gives us a simple analytical expression for time dependence of the mean temperature

$$T = T_m + \left(T_0 - T_m\right) \times \exp\left(-\frac{8\chi}{L^2}t\right), \qquad (7)$$

where $T_0$ is the initial value of the mean temperature of the system.

Fig. 6 shows the comparison of the analytical solution (7) and results of MD simulations with taking the electronic heat conductivity into account. One can see a clear correspondence that supports our assumptions about the determinative influence of the heat supply or withdrawal on the propagation velocity of the phase transition front and about the quasi-stationary character of the local temperature distribution. Our conclusion about the determinative significance of the heat conductivity corresponds to the classical concepts (Stephen problem), but we check its validity for micro-scales and short times. The situation presented in Fig. 6 corresponds to crystallization ($T_0 < T_m$), at which the molten layer becomes thinner and the mean temperature increases. Similar correspondence is obtained in the case of melting ($T_0 > T_m$) as well. This symmetry of melting and crystallization rates means that they both are determined by the same process and the thermal conductivity is the most likely candidate on this role, but not the rate of the material structure changeover, which has no reasons to by symmetric for direct and inverse process.

According to Eq. (7), the mean temperature evolution is determined only by the heat diffusivity $\chi$ and the system size $L$. Therefore, comparison of Eq. (7) with MD results allows one to determine unambiguously the value of $\chi$. In the case of accounting of the electronic heat conductivity, we have $\chi = 3.3 \times 10^{-6} \ \mathrm{m^2/s}$; all curves in Fig. 6 are calculated with using of this value of the heat diffusivity. Comparison between Eq. (7) and MD calculations without electronic heat conductivity gives us $\chi = 2.1 \times 10^{-6} \ \mathrm{m^2/s}$.

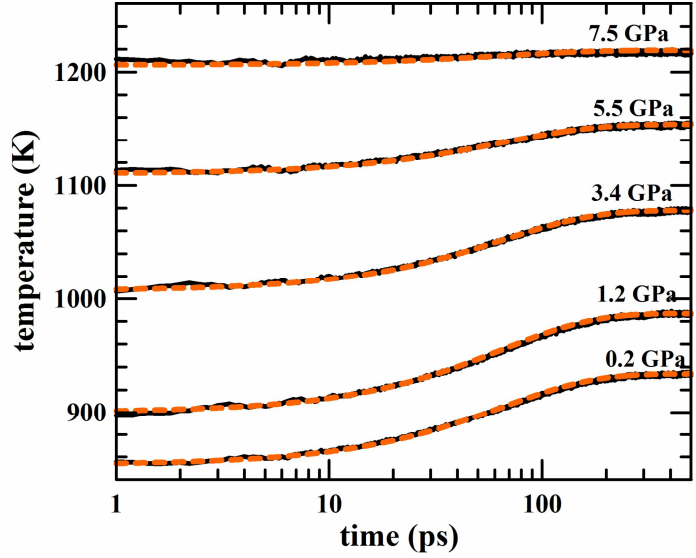

**Fig. 6.** Reaching the phase equilibrium in aluminum with molten layer at various pressures: black solid lines present MD results with accounting of the electronic heat conductivity; dashed orange (light) lines present the analytical solution (7).

Combination of Eq. (7) and Eq. (4) gives us the following analytic expression for the half-thickness of the molten layer

$$R = R_0 + \frac{L}{2}\left(\frac{c_V}{\lambda}\right)(T_0 - T_{\mathrm{m}})\left[1 - \exp\left(-\frac{8\chi}{L^2}t\right)\right], \tag{8}$$

where $R_0$ is the initial value of $R$. The maximal displacement of the interphase front $\Delta R = R\left(t = +\infty\right) - R_0$ at reaching the phase equilibrium in the isolated system is equal to

$$\Delta R = \frac{L}{2}\left(\frac{c_V}{\lambda}\right)\left(T_0 - T_\mathrm{m}\right). \tag{9}$$

Comparison of solution (9) with the displacement observed from MD calculations allows one to determine univocally the ratio $\left(\lambda / c_V\right) = 345\ \mathrm{K}$. Fig. 7 shows the time profile of position of the left crystallization front $L/2 - R$ determined from MD simulations in comparison with the analytical solution (8).

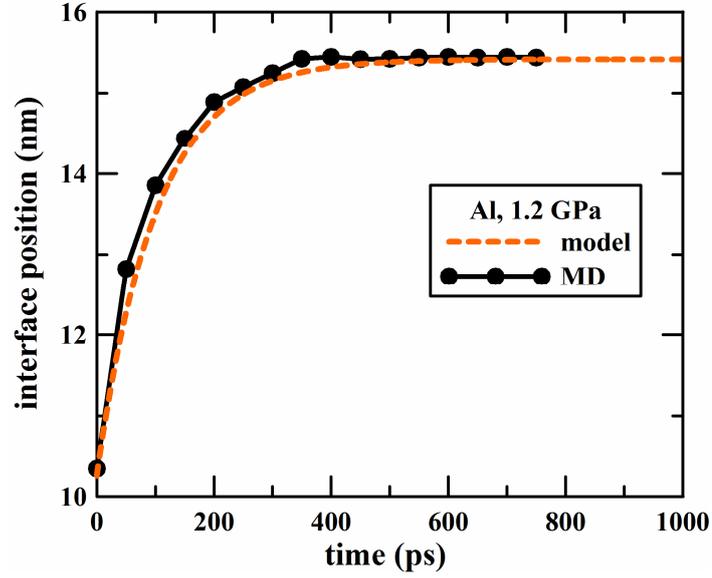

**Fig. 7.** Position of the crystallization front versus time (reaching the phase equilibrium in aluminum at pressure of 1.2 GPa): solid black line with circles presents the MD calculation results; orange (light) dash line presents the analytical solution (8).

### 4.2. Growth rate of spherical melting sites

MD simulation results show that the melting of initially uniform solid goes thorough nucleation and following growth of numerous melting sites. We approximate these sites by spheres for the sake of simplicity. According to Section 4.1, the growth rate of the melting regions is completely determined by the rate of heat supply to the melt interface. The balance between the heat flux and the energy transition into the latent heat of melting in spherical geometry gives Eq. (2) once again, where $R$ means now the radius of the spherical melting region. Consider the spherical region of melt at the constant temperature $T_m$ inside the larger sphere of radius $L/2$; solution of the stationary heat conduction equation gives one the following spatial distribution of local temperature over radius $r$

$$T_{loc}(r) = \begin{cases} T_m, & r \leq R, \\ T_m + R \times (\nabla T)_R - \dfrac{R}{r} \times (\nabla T)_R, & r > R. \end{cases} \quad (10)$$

Averaging the distribution (10) over the system volume $r \in [0, L/2]$ gives us the temperature gradient at the melt interface

$$(\nabla T)_R = \frac{T - T_m}{R} \left\{ 1 - \alpha - \frac{3}{2}\alpha^{1/3} + \frac{3}{2}\alpha^{2/3} \right\}^{-1}, \quad (11)$$

where $T$ is the mean temperature in the system; here we use the fact that the volume fraction of melt is equal to the ratio $\alpha = (2R/L)^3$. Combining Eq. (2) and Eq. (11), we get the following equation of the melting sites growth

$$\frac{dR}{dt} = \left( \frac{c_V}{\lambda} \right) \chi \left( \frac{T - T_m}{R} \right) \left\{ 1 - \alpha - \frac{3}{2}\alpha^{1/3} + \frac{3}{2}\alpha^{2/3} \right\}^{-1}. \quad (12)$$

### 4.3. Nucleation rate of melting sites

A homogeneous melting mode takes place at fast temperature increase; the phase transition starts from nucleation of a number of melting sites in this mode. The nucleation of sites can be described using the thermofluctuational approach. Suppose that the minimal stable melting site contains $N_{cr}$ atoms. The difference between the energy of a molten element containing $N_{cr}$ atoms at melting temperature $T_m$ and the energy of an overheated element containing the same number of atoms at temperature $T > T_m$ is equal to $E_{cr} = N_{cr} \times m_1 \left[ \lambda - c_V (T - T_m) \right]$, where $m_1$ is the mass of one atom. Nucleation of melting sites due to thermal fluctuations is a stochastic process. The total energy of a system with a melting site is $E_1 = E_0 + E_{cr}$, where $E_0$ is the total energy of similar system without melting site. In accordance with Gibbs distribution [25], the probability of a system state with the energy $E$ is equal to $A \times \exp\left(-E/\left(k_B T\right)\right)$, where $k_B$ is the Boltzmann constant, $A$ is the normalization constant that does not depend on $E$. Particularly, this expression gives the probability of the system state with a certain position of the melting site, if we write $E_1$ instead of $E$. The melting sites can arise in different positions; the number of possible positions can be estimated as the ratio of the total number $N$ of atoms in the considered system to the number $N_{cr}$ of atoms inside one melting site. Therefore, the probability of the system state with arbitrary position of site is equal to $A \times N \times \exp\left(-E_1/\left(k_B T\right)\right)$, while the probability of the initial uniform state is equal to $A \times \exp\left(-E_0/\left(k_B T\right)\right)$. The ratio of these probabilities, $N \times \exp\left(-E_{cr}/\left(k_B T\right)\right)$, gives the probability of the system transition from uniform state to the state with a melting site, it means, the nucleation of one melting site in arbitrary position. A characteristic time of change of state due to thermal vibrations is $R_{cr}/c_s$, где $R_{cr} = d\left[3N_{cr}/\left(4\pi\right)\right]^{1/3}$ is the radius of the melting site, $d = \left(m_1/\rho\right)^{1/3}$ is the average interatomic distance, $\rho$ is the density of solid phase, and $c_s$ is the sound speed. Thus, the rate of homogeneous nucleation of melting sites per unit volume can be written as following

$$\frac{dn}{dt} = \left(\frac{4\pi}{3}\right)^{1/3} \frac{c_s}{d^4} \frac{1}{N_{cr}^{4/3}} \times \exp\left(-N_{cr} \times m_1 \frac{\lambda - c_v\left(T - T_m\right)}{k_B T}\right), \tag{13}$$

where $n$ is the melting sites concentration; $k_B$ is Boltzmann constant.

The minimal number $N_{cr}$ of atoms in a stable melting site is interpreted by us as a parameter of the continuum model; value of this parameter for aluminum $N_{cr} = 45$ is obtained from comparison with MD data (see Section 4.5). A qualitative interpretation of this value can be obtained from comparison between the typical hydrodynamic time $\tau_h = R_{cr}/c_s$ and the typical time $\tau_t = \left(R_{cr}\right)^2/\chi$ of heat conductivity. The first typical time $\tau_h$ determines the rate of mechanical unloading (expansion) of substance within an arising site, which experiences the phase transitions and increases the specific volume; this unloading is necessary for the phase transition completion. The second typical time $\tau_t$ defines the dissipation rate of the thermal energy from the fluctuation area to the surrounding substance, which temporally becomes colder than the fluctuation area. The time $\tau_t$ of thermal conductivity is less than the time $\tau_h$ of mechanical unloading for small fluctuation areas; the energy dissipates from the potential melting site before the phase transition completion as a result. Therefore, the stable melting site corresponds to the condition $\tau_t \geq \tau_h$, which gives the following estimation

$$N_{cr} = \frac{4\pi}{3}\left(\frac{\chi}{c_s d}\right)^3. \tag{14}$$

The estimation (14) corresponds to the value $N_{cr} \approx 45$ obtained from comparison with MD at the thermal diffusivity equal to $\chi = 2.6 \times 10^{-6}$ m$^2$/s, which is close to the value obtained for the ionic thermal conductivity alone (see Section 4.1). This is understandable because the site formation time $\tau_t = \tau_h \approx 0.1$ ps is considerably less than the time of electron-ion relaxation, which is about tens of picoseconds.

### 4.4. Equations system of continuum model

The continuum model of melting proposed in previous Subsections can be used for simulation of non-equilibrium melting of substance in strong shock waves or in the energy absorption area of the high-current electron or powerful ion beams or under the exposure to intensive laser irradiation. For this purpose we include the melting model as a component into the standard continuum mechanics system

$$\frac{\mathrm{d}\rho}{\mathrm{d}t} = -\rho\left(\nabla \cdot \mathbf{v}\right), \tag{15}$$

$$\frac{\mathrm{d}\mathbf{v}}{\mathrm{d}t} = \frac{1}{\rho}\left[-\left(\nabla P\right) + \left(\nabla \cdot \mathbf{S}\right)\right], \tag{16}$$

$$\frac{dU}{dt} = \frac{1}{\rho}\left[-P(\nabla \cdot \mathbf{v}) + (\mathbf{S} : \mathbf{w})\right] + D - \frac{\lambda}{\rho}\frac{d\alpha}{dt}, \tag{17}$$

where Eq. (15) is the equation of continuity, Eq. (16) is the equation of motion, Eq. (17) is the equation for internal energy; $\rho$ is the density; $\mathbf{v}$ is the velocity field; $P$ is the pressure; $\mathbf{S}$ is the tensor of stress deviators; $\mathbf{w}$ is the tensor of plastic deformation; $D$ is the energy absorption function that takes into account the effect of irradiation; del $\nabla$ means the vector of spatial derivatives. In Eq. (17), $U$ is the part of specific internal energy, which does not include the latent heat of melting. This part corresponds to the internal energy definition used in the wide-range equations of state without explicit accounting of the solid-liquid phase transition, [26] for instance. The last term in the right-hand part of Eq. (17) takes into account the latent heat of melting in this case. The equation of state determines the functional dependences $P = P(U, \rho)$, $T = T(U, \rho)$. Lagrange frame of reference is used in Eqs. (15)-(17). The tensor of stress deviators $\mathbf{S}$ and the tensor of plastic deformation $\mathbf{w}$ is calculated using the dislocation plasticity model [27,28] in the solid metal, and they are zero in the melt. The volume faction $\alpha$ of the molten substance can be calculated as

$$\alpha = \frac{4\pi}{3}R^3 n. \tag{18}$$

Eqs. (15)-(18) are complemented by the equations for concentration $n$ and mean radius $R$ of the melting sites, which are rewritten in the following form

$$\frac{dn}{dt} = \left(\frac{4\pi}{3}\right)^{1/3}\frac{c_s}{d^4}\frac{1}{N_{cr}^{4/3}} \times \exp\left(-N_{cr} \times (c_V m_1)\frac{(\lambda/c_V)-(T-T_m)}{k_B T}\right) \times (1-a), \tag{19}$$

$$\frac{dR}{dt} = \left(\frac{c_V}{\lambda}\right)\chi\left(\frac{T-T_m}{R}\right)\left\{1-\alpha-\frac{3}{2}\alpha^{1/3}+\frac{3}{2}\alpha^{2/3}\right\}^{-1} - \frac{R^3 - R_{cr}^3}{3R^2}\cdot\frac{1}{n}\frac{dn}{dt}. \tag{20}$$

$R$ and $n$ are defined in each point of substance, and, thus, form the fields. Instantaneous spontaneous melting of metal occurs at temperatures $T \geq T_m + \lambda/c_V$, while a potential barrier exists at lower temperature. The barrier is connected with the latent heat of melting $\lambda$ and can be overcome at the expense of thermal fluctuation. This process is described by Eq. (19), where the last multiplier in the right-hand part accounts that the nucleation of sites takes place only inside the solid phase. Motion of phase boundaries connected with the growth of previously arisen melting sites is a competing process described by Eq. (20). We do not take into account the size distribution of the melting sites; therefore, an additional in comparison with Eq. (12) term is added into the right-hand side of Eq. (20), which accounts an effective change of the mean radius due to formation of new sites with smaller radius $R_{cr}$.

The equation system is solved numerically. Continuum mechanics equations (15)-(17) are solved by the finite-difference method proposed in Ref. [29]. Equations (19), (20) are time-integrated using the explicit Euler scheme, the solution stability is provided by using the small enough time step. The numerical solution was realized as an expansion of CRS computer code [30]. At modeling the high-current electron beam action, the energy absorption function entered Eq. (17) is calculated from the electron transport problem solution with using the method proposed in Ref. [31]. We use the melting model parameters, which are defined from comparison with MD simulation results and collected in Table 1; the melting curve (1) is also used. Pressure $P$, temperature $T$ and specific heat capacity $c_V$ are calculated from internal energy $U$ and density $\rho$ with using the wide-range equation of state [26].

Tabel 1. Parameters of the continuum model of melting for aluminum

| Parameter | Value |
|---|---|
| $\chi \left[ \mathrm{m^2/s} \right]$ | $3.3 \times 10^{-6}$ |
| $\left( \lambda / c_V \right) \left[ \mathrm{K} \right]$ | 345 |
| $N_{cr}$ | 45 |

### 4.5. Comparison of continuum model and MD data

Fig. 8 shows the continuum model results in comparison with MD data for the problem of heating an initially uniform monocrystal of aluminum with the constant energy release power $D$. Eqs. (15), (16) are not solved at continuum modeling in this simplified case; the volume growth rate $(\nabla \cdot \mathbf{v})$ is chosen from the condition of maintaining zero pressure $P$, which corresponds to MD. According to both MD and continuum model, a monotonous increase of temperature takes place in the initial stage; aluminum remains solid and passes into a metastable overheated state. The maximal overheating reaches 170-220 K depending on $D$. In some temperature range, which is varied for different heating rates, a massive nucleation of melting sites begins with their following growth that initially reveals itself in some deceleration of the temperature increase; thereafter the temperature drop takes place due to the accumulated thermal energy transfer into the latent heat of melting. During the time interval corresponding to the negative slope of the temperature-time curve, the melt volume fraction rapidly increases from an almost zero value to unit, which corresponds to complete melting. The heating of uniform melt goes further with a monotonic temperature increase with time.

Comparison of MD data and continuum model results allow us to determine the value $N_{cr} = 45$ for the parameter controlling the melting sites nucleation (see Eq. (19)). A satisfactory agreement between MD and continuum calculations is observed for all considered heating rates at this value of parameter (Fig. 8). Continuum model overestimates (within 8%) the overheating of solid aluminum, as well as the value of temperature drop in the process of rapid melting. The difference decreases together with the heating rate decrease. As it follows from comparison of curves, the growth rate of large sites (the final part of the energy drop) is described within the continuum model worse than that for small sites (the initial part of the energy drop). It can be explained by approximations made at the derivation of Eq. (12).

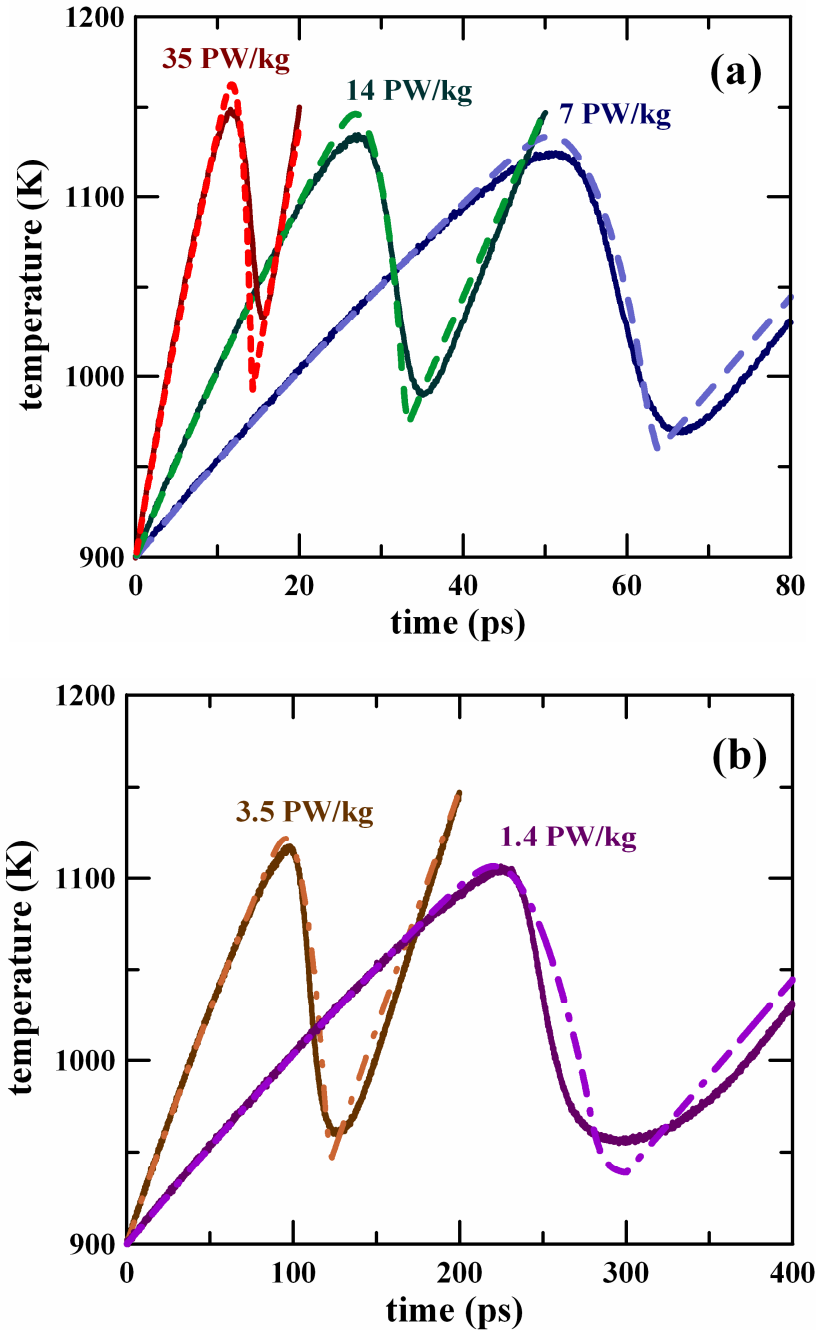

**Fig. 8.** Melting of aluminum with the constant power of energy deposition ((a) – 35, 14 and 7 PW/kg; (b) – 3.5 and 1.4 PW/kg): results of MD simulations with accounting the electronic heat conductivity (solid lines) and continuum model calculations (dash lines).

## 5. Melting of aluminum inside the energy absorption area of an electron beam

As an example of application the developed continuum model of melting, we consider the action on aluminum of the high-current electron beam with parameters corresponding to SINUS-7 electron accelerator (Tomsk, Russia, Institute of High Current Electronics SB RAS) [5]. The maximal energy of electrons is 1.35 MeV, the pulse duration is 45 ns, the maximal power density is 34 GW/cm$^2$, and the incident energy density is 1.2 kJ/cm$^2$. Calculations are performed in 1D with using the CRS computer code [30] and the wide-range equation of state [26]. Experimental oscillograms of voltage and current density are used.

Fig. 9 presents the spatial distributions of the substance temperature $T$, the melting temperature $T_m$ depending on the local pressure (see Eq. (1)) and the volume fraction $(1 - \alpha)$ of solid phase for time moments 25, 30 and 35 ns from the beginning of irradiation. The considered electron beam heats the aluminum layer about 2 mm in thickness; spatial distribution of energy release power is non-uniform that leads to non-uniform temperature distributions. The energy release power reaches 0.08 PW/kg in maximum that leads to the heating rate of about 70 K/ns. A complete melting occurs in the central part of the energy absorption area to the moment of time of 25 ns (Fig. 9(a)). Temperature of uniform melt begins to exceed $T_m$ in this central area, while it is close to $T_m$ at the edges of the complete melting zone. Two regions of partial melting adjoin the complete melting zone from both sides. $T \approx T_m$ in the internal parts of these partial melting regions that means the reaching the phase equilibrium; increase of the melt volume fraction goes here to the extent of energy supply by the beam. Aluminum is overheated $T > T_m$ within the external parts of the partial melting zones; the melt volume fraction goes here at the expanse of both the energy transfer from by beam and the leaving the metastable overheated state. Areas of overheated solid metal adjoin the partial melting zones. The overheating reaches $T - T_m \approx 100$ K. The left zone of partial melting disappears to the time moment of 30 ns; aluminum becomes completely melted up to the exposed free surface (Fig. 9(b)). The right zone of partial melting moves deep into the target in the course of time (Fig. 9(b),(c)); the molten layer thickness increases till the beam action termination.

Thus, the proposed continuum model allows one to describe in detail the non-equilibrium solid-liquid phase transition within the energy absorption area of a high-current electron beam.

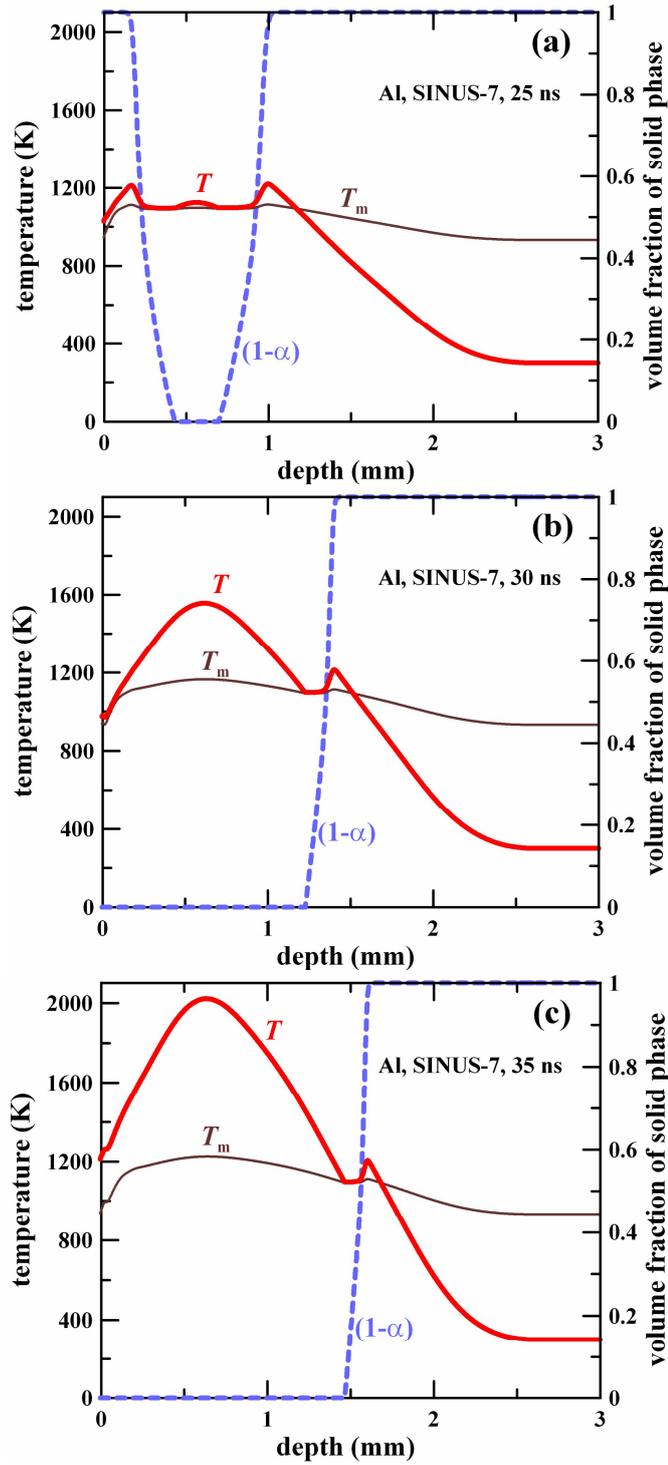

**Fig. 9.** Spatial distributions of substance temperature $T$ (thick solid lines), melting temperature $T_{\mathrm{m}}$ (thin solid lines) and volume fraction $(1-\alpha)$ of solid phase (dash lines) in consequent moments of time: irradiation of aluminum by the high-current electron beam with parameters of SINUS-7 accelerator. Initial target thickness is 3 mm; the electrons of beam fall normally on the left surface of target.

## 6. Conclusions

MD investigation of the non-equilibrium phase transition of aluminum is performed in frames of two problem statements: (i) the reaching of equilibrium in an isolated system with a flat interface between solid and liquid; (ii) the melting of an initially homogeneous monocrystal heated by a source with constant power of energy release. Analysis of the obtained results shows that the changeover of the material structure requires a negligible time during the melting or crystallization front propagation, while the propagation velocity is completely determined by the rate of the energy supply or withdrawal from the surrounding substance. MD simulations are performed both with and without accounting of the electronic subsystem; the electrons increase the heat diffusivity in 1.5 times only in the considered situation that can be explained by the comparability of the typical times of the melting processes investigated in MD with the characteristic time of electron-ion relaxation, which is about 10 ps. Accounting of electrons almost not influence on the solid phase overheating reached at the aluminum heating with the constant energy release rate.

A continuum model of melting is purposed, in which the current phase state is described in terms of fields of concentration and size of melting sites. Growth equation for melting areas is derived from the heat fluxes analysis. Nucleation equation for melting sites is formulated basing on the thermofluctuational approach. The method of determination of the model coefficients with using the MD simulation results is purposed. The continuum model is applied to the problem of the non-equilibrium melting within the energy absorption area of the high-current electron beam.

### Acknowledgments

The work was supported by the grant from the Russian Foundation for Basic Research (Project No. 15-32-21039).